\documentclass[journal,a4paper,onecolumn,draftcls,12pt]{IEEEtran}

\IEEEoverridecommandlockouts

\newcommand\hmmax{0}
\newcommand\bmmax{0}

\usepackage{graphicx}
\usepackage{tikz}
\usepackage{pgfplots}
\usepackage{xcolor}
\usepackage{color, colortbl}
\usepackage{amsmath}
\usepackage{bbm}
\usepackage{amsfonts, amssymb}
\usepackage{mathtools}
\usepackage{cite}
\usepackage[nolist]{acronym}

\usepackage{booktabs} 		
\usepackage{diagbox}		
\usepackage{upgreek}
\usepackage{bbm}
\usepackage{Files/bm}
\usepackage{Files/winsnotation}
\usepackage{Files/LVcolours} 

\usepackage[ruled,vlined,noresetcount]{algorithm2e}
\usepackage{setspace}	 	
\usepackage{comment}
\usepackage{multirow}

\usepackage{array}
\newcolumntype{C}[1]{>{\centering\arraybackslash}p{#1}}

\makeatletter
\def\endthebibliography{%
  \def\@noitemerr{\@latex@warning{Empty `thebibliography' environment}}%
  \endlist
}
\makeatother
\usepackage{amsthm}
\theoremstyle{definition}

\pgfplotsset{compat=1.17}
\usepgfplotslibrary{fillbetween}

\begin{document}

\title{An Implementation of the Optimal Scheme \\ for Energy Efficient Bus Encoding}

\author{Lorenzo~Valentini,~\IEEEmembership{Member,~IEEE,}
        and~Marco~Chiani,~\IEEEmembership{Fellow,~IEEE}
\thanks{The authors are with the Department of Electrical, Electronic, and Information Engineering ``Guglielmo Marconi'' and CNIT/WiLab, University of Bologna, 40136 Bologna, Italy. E-mail: \{lorenzo.valentini13, marco.chiani\}@unibo.it.}
}

\maketitle 
\markboth{}{Valentini, Chiani: Analysis and Design of Energy-Efficient Bus Encoding}

\begin{acronym}
\small
\acro{i.i.d.}{independent and identically distributed}
\acro{BSC}{binary symmetric channel}
\acro{CNN}{convolutional neural network}
\acro{ECC}{error correcting code}
\acro{LUT}{look-up-table}
\acro{i.i.d.}{independent and identically distributed}
\acro{MPPM}{multi pulse-position modulation}
\acro{MUX}{multiplexer}
\acro{NVM}{non-volatile memories}
\acro{NoC}{Network-on-Chip}
\acro{PI}{Partitioned Inversion}
\acro{PRAM}{phase-change random access memory}
\acro{PPM}{pulse-position modulation}
\acro{PR}{Pure Random}
\acro{RI}{Random Inversion}
\acro{r.v.}{random variable}
\acro{DBI}{Data Bus Inversion}
\acro{DMS}{discrete memoryless source}
\acro{SI}{Shift Inversion}
\acro{SSO}{simultaneous switching output}
\acro{WEM}{Write-Efficient Memory}
\end{acronym}
\setcounter{page}{1}

\begin{abstract}
In computer system buses, most of the energy is spent to change the voltage of each line from high to low or vice versa. 
Bus encoding schemes aim to improve energy efficiency by 
limiting the number of transitions between successive uses of the bus. 
We propose an implementation of the optimal code with reduced  number of clock cycles. 
\end{abstract}

\section{Introduction}

The efficient use of buses represents an important topic for low-power computer systems design. For example, it is estimated for \ac{NoC} applications that the dynamic power dissipated by interconnections is over $50\%$ of the total dynamic power \cite{Mag:04,Wol:08,Jaf:14,Vel:20}. 
In addition to their conventional application in memory interfaces \cite{Connor22:BusEncodingPAMMemory}, with the increasing number of artificial intelligence applications, the efficient use of buses is also becoming an important challenge in the implementation of \acp{CNN}, where systolic array architectures are used to perform matrix multiplications \cite{peltekis2023low}.
For a bus consisting of several lines used to transport binary tuples, 
most of the energy can be attributed to  the change of status, from a logical $0$ to a $1$ or vice versa, of each line, so that the energy cost is mainly related to the Hamming distance between consecutive tuples on the bus \cite{Fle:87,Tab:90,Sta:95,Sta:97}. 
Without bus encoding, to send a binary $k$-tuple (information) we need a $k$-lines bus. With bus encoding, the information $k$-tuple is mapped into a binary $n$-tuple (codeword), $n>k$, then transmitted over an $n$-lines bus.  The objective is to consume less energy, at the cost of $b=n-k$ additional bus lines, plus some signal processing. For this reason, the mapping and the codebook are designed to minimize the average distance of a codeword from the previously transmitted codeword, i.e., the weight of their element-wise XOR.  Bus encoding is therefore a line coding technique, with low-weight codewords used in a differential manner.  
The design of low-weight line codes finds application in several systems, including \acp{NoC}, dynamic random access memories (DRAM), and peripheral component interconnect (PCI) \cite{Ahl:89,BenMic:00,Mag:04,Wol:08,Yua:08,Jaf:14,Vel:20,Gho:18,Wan:20}. 
The \ac{PPM} technique used in optical communication systems is another example of the application of low-weight codes \cite{GagKar:76,ccsds:19a,ModLocVal:22}. 

 
In the context of bus encoding, several techniques have been proposed to reduce the number of transitions between successive uses of the bus, starting from the seminal works in \cite{Fle:87,Tab:90,Sta:95,Nat:04,Ben:98}. 
Some of these works assume a memoryless source, emitting random binary tuples to be transported over the bus \cite{Fle:87,Tab:90,Par:92,Sta:95,Nak:96,Sta:97,Kom:99,Nat:04,Ala:15,Kan:22DBI,Che:2022TabBusEncoding}, others focus on sources with memory  \cite{Ben:97,Ben:98,Mus:98,Ram:99,Ram:99b,Ben:00,For:00,Tie:03,Sun19:AdaptiveBusEncoding}.
The formulation of sub-optimal schemes provides valuable flexibility when optimal solutions \cite{Tab:90, Ram:99} are not practically achievable. In such instances, a trade-off in performance is often accepted to facilitate implementation. 

In this note we discuss an implementation for the optimal scheme. 


\section{General Framework}
\label{sec:preliminary}

\subsection{Problem Formulation}\label{subsec:GenFrame}

\begin{figure*}[t]
    \centering
    \includegraphics[width=\textwidth]{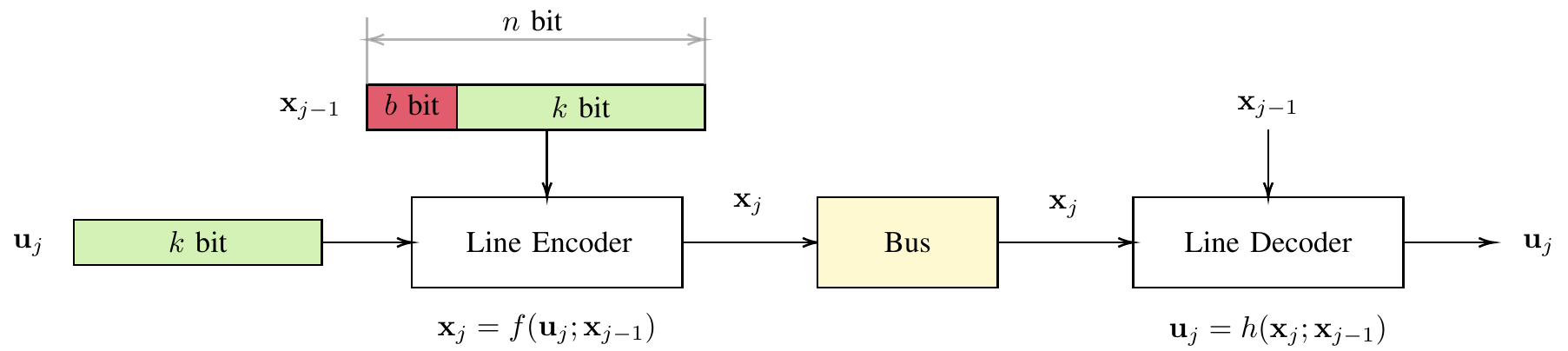}
    \caption{General Framework.}
    \label{fig:GenFrame}
\end{figure*}

Let us consider a \ac{DMS} emitting information binary $k$-tuples $\mathbf{u}_1, \mathbf{u}_2, \ldots, $ where $\mathbf{u}_j$ is the binary $k$-tuple at time $j$. The $\mathbf{u}_j$ are assumed uniformly distributed over the alphabet $\{ 0, 1\}^k$. As represented in Fig.~\ref{fig:GenFrame}, the $\mathbf{u}_j$ is mapped by a line encoder into the codeword $\mathbf{x}_j \in \{ 0, 1\}^n$, whose length is $n = k + b$, and $b$ represents the number of redundancy bits (added lines). 
Hence, the problem is to find a function $\mathbf{x}_{j} = f(\mathbf{u}_{j}; \mathbf{x}_{j-1})$, minimizing the average distance between successive codewords
\begin{align}
\label{eq:DGenFrame}
    D = \E{ d_\mathrm{H}(\mathbf{x}_{j},\,\mathbf{x}_{j-1})} = \E{w(\mathbf{x}_{j} \oplus \mathbf{x}_{j-1})}
\end{align}
where $d_\mathrm{H}(\cdot, \cdot)$ is the Hamming distance, $w(\cdot)$ is the weight function, and the expectation is performed over all possible $\mathbf{x}_{j-1}$ and $\mathbf{u}_{j}$.
The mapping $f(\cdot ; \cdot)$ must be invertible given $\mathbf{x}_{j-1}$, so that  $\mathbf{u}_{j}$ can be recovered given $\mathbf{x}_{j}$ and $\mathbf{x}_{j-1}$.
From \eqref{eq:DGenFrame}, we see that the problem of minimizing the average distance between consecutive codewords is equivalent to minimizing the average weight of the differential codewords $\mathbf{d}_j = \mathbf{x}_{j} \oplus \mathbf{x}_{j-1}$.
%

Let us start with the uncoded case, where $\mathbf{x}_j = \mathbf{u}_j$, $b=0$. Here, we observe that, given two tuples of $k$ bits generated uniformly at random, the probability that they are at distance $d=0, \dots, k$ is
\begin{align}
\label{eq:PdistGivenk}
   p(d) = \Prob{d_\mathrm{H}(\mathbf{u}_{j},\,\mathbf{u}_{j-1})=d}=\binom{k}{d}\,\frac{1}{2^k}
\end{align}
and the average distance is, as expected, 
\begin{align}
    D_\mathrm{unc}(k) = \sum_{d = 0}^{k}\binom{k}{d}\,\frac{d}{2^k} = \frac{k}{2}\,.
\end{align}

\subsection{Minimum Achievable Average Distance}\label{subsec:Bound}
Let us analyze the minimum average distance \eqref{eq:DGenFrame} for arbitrary $n,k$. To describe the optimal encoder we observe that we have $2^k$ equiprobable  $k$-tuples to be mapped into codewords of $n$ bits. 
Since we want to minimize $D = \E{w(\mathbf{d}_{j})}$, the optimal line encoder is the one that first maps $\mathbf{u}_j$ into $\mathbf{d}_j \in \mathcal{D}$, where the codebook $\mathcal{D}$ contains the $2^k$ lowest weight $n$-tuples.
This is due to the fact that the weight of $\mathbf{d}_j$ represents the number of transitions which will occur in the bus line. 
Hence, if we construct a codebook for $\mathbf{d}_j$ using all possible sequence with the lowest weight, we obtain the optimal coding strategy for this problem.
The optimal codebook $\mathcal{D}$ is thus composed of $\binom{n}{0}$ zero-weight codewords, $\binom{n}{1}$ one-weight codewords, $\binom{n}{2}$ two-weight codewords, and so on. 
Then, the line codewords are obtained by differential encoding as $\mathbf{x}_{j} = \mathbf{d}_{j} \oplus \mathbf{x}_{j-1}$. This code construction can be found in \cite{Ahl:89,Tab:90}.
%
%
%
The average distance of the optimal $(n,k)$ line code is \cite{Ahl:89}
\begin{align}
\label{eq:Dopt}
    D_\mathrm{opt}(k, b) = d_\mathrm{max} - \sum_{i = 0}^{d_\mathrm{max} - 1} \frac{d_\mathrm{max} - i}{2^k}\,\binom{n}{i}\,.
\end{align}
where $d_\mathrm{max}$ is the minimum integer such that 
$\sum_{i=0}^{d_\mathrm{max}}\binom{n}{i} \ge 2^k$.
%

\section{Optimal Code Implementations} 
\label{sec:Opt}
For the optimal scheme presented in the previous section, the critical issue is the feasibility of encoders and decoders. 
In fact, assuming a differential scheme where $\mathbf{x}_{j} = \mathbf{x}_{j-1} \oplus \mathbf{d}_{j}$, the optimal encoder must injectively associate a low-weight codeword to each binary $k$-tuple, i.e., it has to implement the mapping $\{0,1\}^k \to \mathcal{D}$.
For a generic number of added lines $b$ and small $k$ this mapping can be implemented through a \ac{LUT} with $2^k$ entries \cite{Tab:90}. However, this approach is not feasible for bus sizes of practical interest. 
In the following, we present other possible implementations.

\subsection{Minimum Redundancy Case: Data Bus Inversion}\label{subsec:Singleinv}
Assume we have just one additional line, $b=1$, so that the bus has $n=k+1$ lines. 
In this particular case, the optimal scheme has been derived in \cite{Fle:87,Sta:95} and is known as \ac{DBI}. 
 With \ac{DBI}, it is transmitted $\mathbf{u}_{j}$ concatenated with a $0$, or its complement $\overline{\mathbf{u}}_{j}$ concatenated with a $1$, depending on which one is closest to the previous codeword on the bus. The additional line serves therefore to indicate if the inversion has been applied or not. 


\subsection{Maximum Redundancy Case: PPM$_0$}\label{subsec:maxred}
For a given information size $k$, the minimum value for the average distance is obtained when the differential codebook $\mathcal{D}$ contains only weight zero and weight one $n$-tuples. This requires a bus size $n=2^k-1$, and a number of added lines $b=2^k-1-k$. The scheme is the most effective as it achieves the smallest possible average distance, but it is also the most redundant one. Its average distance, which is the smallest possible for a given $k$, is 
\begin{align}
\label{eq:DPPM}
    D_{\min}(k) = D_\mathrm{opt}(k, 2^k-1-k)=1 - \frac{1}{2^k} \,. 
\end{align}
This approach reminds \ac{PPM}, where exactly one pulse is used over $n$ slots to send information. \ac{PPM} is widely adopted, e.g., in optical communication systems \cite{GagKar:76,ccsds:19a,ModLocVal:22}.
More precisely, the maximum redundancy scheme can be interpreted as a \ac{PPM} mapping with the addition of the all-zero codeword, that will be denoted as  PPM$_0$.

\subsection{Implementation based on syndrome decoders (coset codes)}\label{subsec:syndromeLike}

\begin{figure*}[t]
    \centering
    \includegraphics[width=\textwidth]{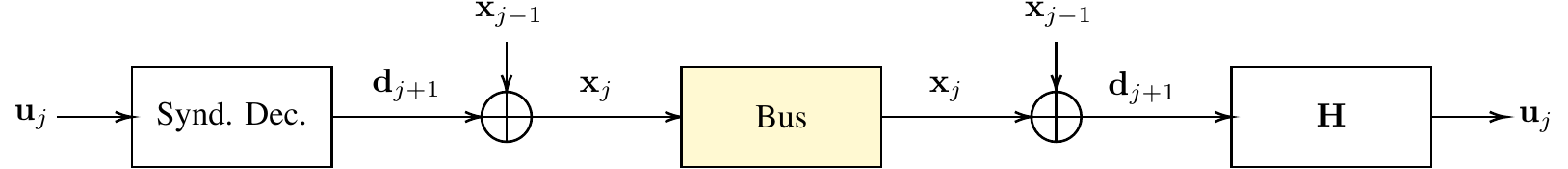}
    \caption{Syndrome-based low-weight codes for bus encoding.}
    \label{fig:SyndBased}
\end{figure*}

The implementation for low-weight codes can be based on coset codes, i.e., based on \acp{ECC} designed for the \ac{BSC} \cite{Ahl:89,Coh:90,Fan:00,Jac:13}. 
We recall that, for a $(N,K)$ binary linear \ac{ECC} each codeword $\V{c} \in \{ 0, 1\}^{N}$ must fulfill
\begin{align}
    \M{H} \, \V{c} = \V{0}
\end{align}
where $\M{H} \in \{ 0, 1\}^{(N-K) \times N}$ is the code parity check matrix and all operations are computed in the binary field. Then, for any received vector $\V{r} \in \{ 0, 1\}^{N}$ we can write
\begin{align}\label{eq:Hr}
    \M{H} \, \V{r} = \M{H} \, (\V{c} + \V{e}) = \M{H} \, \V{e} = \V{s}
\end{align}
where $\V{e} \in \{ 0, 1\}^{N}$ is the error pattern introduced by the \ac{BSC}  and $\V{s} \in \{ 0, 1\}^{(N-K)}$ is the associated error syndrome. 
A decoder based on the minimum Hamming distance criterion between the observed  $\V{r}$ and all possible $\V{c}$ can thus be implemented by looking for the $\V{e}$ of lowest weight satisfying \eqref{eq:Hr}.  
For some syndromes the lowest weight error pattern will be unique, for some others there could be more possible patterns of lowest weight compatible with the syndrome. 
The latter case will correspond to error weights greater than $(d_{\min} -1)/2$, where $d_{\min}$ is the minimum Hamming distance of the code \cite{RyaLin:09}. 
For example, for codes like Hamming and Golay, algebraic decoders can be used to associate, to each syndrome, patterns of no more than $(d_{\min} -1)/2$ errors. 
In general we need complete decoders, i.e., decoders which produce a low weight error patter for each possible syndrome \cite{RyaLin:09}. 
With this interpretation, a minimum distance decoder which associates a low (ideally the lowest) weight error pattern to every possible syndrome can be seen as an encoder which associates to every possible $(N-K)$-tuple $\V{s} \in \{ 0, 1\}^{(N-K)}$ a low weight $N$-tuple $\V{e}\in \{ 0, 1\}^{N}$. The bus encoder obtained by setting $\V{s} = \V{u}_j$ and $\V{e} = \V{d}_{j}$ is depicted in Fig.~\ref{fig:SyndBased}. In this scheme, from an $(N,K)$ linear code 
we can construct a low-weight line encoder with bus length size $n=N$ and information bus size $k = N-K$. 
If we use a linear code with maximum correction capability $t$ the resulting scheme is therefore an implementation of the optimal bus encoder with codewords of maximum weight $t$. 
For example, let us consider the $(N,1)$ repetition codes, $N$ odd, $t=(N-1)/2$; the Hamming $(2^m-1, 2^m-1-m)$, $t=1$ codes; the $(23, 12)$, $t=3$ Golay code \cite{RyaLin:09,Eli:87,WeiWei:90}. 

For the $(N,1)$ repetition code we can encode $k= N-1$ bits into $n=N$ lines, and the resulting scheme is equivalent to the \ac{DBI}. 

On the opposite side, for the Hamming decoder the maximum weights of the output error patterns is $t=1$, so it implements the maximum redundancy encoder of Section~\ref{subsec:maxred}. For example, with the Hamming $(15,11)$ decoder we will map $k=N-K=4$ bits on a bus with $n=15$ lines, achieving the minimum average distance $D_{\min}(4)=15/16$, to be compared with the uncoded case average distance $D_\mathrm{unc}(4)=2$. With respect to the uncoded case, the reduction in terms of average number of transitions obtained with bus encoding is therefore $D_{\min}(4)/D_\mathrm{unc}(4)=15/32 \simeq 0.46875$, which is the lowest possible for $k=4$. So, the number of transitions is more than halved, but at the price of a large number of added lines. 

A less redundant scheme can be obtained from the Golay $(23, 12)$, $t=3$ code, where a syndrome decoder \cite{Eli:87,WeiWei:90} maps $23-12=11$ bits into a bus with $23$ lines, with maximum weight $3$. This is a way to implement the optimal bus encoder with $k=11$ and $n=23$. From \eqref{eq:Dopt} we get $D_\mathrm{opt}(11, 12)=2921/1024$, so that the transitions reduction is $D_\mathrm{opt}(11, 12)/D_\mathrm{unc}(11)=0.5186$.
In this case, by approximately doubling the number of lines we approximately halve the number of transitions. Of course this approach can be used for larger $k$, by partitioning the information in blocks of $11$ bits that are mapped to $23$ lines. 

More in general, minimum distance decoders (coset decoders) can be used to realize the optimal line encoder for other dimensions and redundancy, although the implementation is challenging for large $k$ and $b>1$. 

\subsection{Implementation Based on Combinatorial Number System}
\label{subsec:GeneralOptimalScheme}

\begin{figure*}[t]
    \centering
    \includegraphics[width = \textwidth]{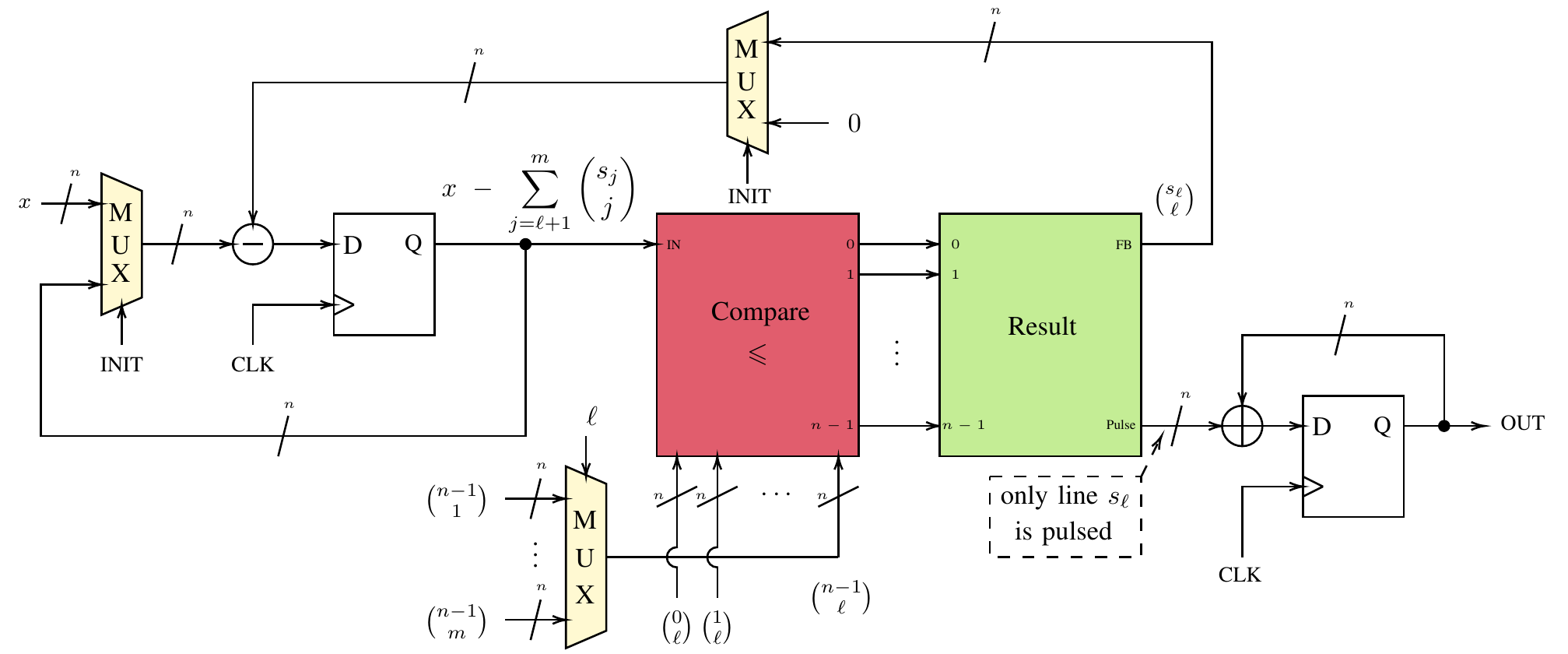}
    \caption{Implementation of a \ac{MPPM} modulator using pre-allocated binomial coefficients. The scheme is able to compute $\text{MPPM}\left(x; \,m\right)$ in $m$ clocks.}
    \label{fig:OptSchemeImpl}
\end{figure*}

The generalization of \ac{PPM}, discussed in Section~\ref{subsec:maxred}, is known as \ac{MPPM}, where exactly $m$ pulses are activated in $n$ slots. 
%
The optimal mapping $\{0,1\}^k \to \mathcal{D}$ described in Section~\ref{subsec:Bound} can thus be interpreted as 
$d_\mathrm{max}+1$ \ac{MPPM} modulators, each of them using a different number of pulses, from $0$ to $d_\mathrm{max}$.
An implementation of the optimal encoder using the combinatorial number system was discussed for the first time in \cite{Ram:99}.
We now describe a variant implemented with \ac{MPPM} modulators. The main difference is that, while the scheme in \cite{Ram:99} evaluates each bit of the $n$-length sequence to find out if it is a ``0'' or a ``1'', our solution produces the positions where to put a ``1''. So, our variant requires $m$ clocks, while the scheme in \cite{Ram:99} requires $n$ clocks (note that $n\ge2m$ is always verified for $b>0$).

To describe our variant, let us denote with $u\in \mathbb{N}$ the integer representation of $\mathbf{u}_j$. First, we compute the weight of the codeword associated with $\mathbf{u}_j$ as the minimum integer $m$ such that
\begin{align}
\label{eq:findmpulses}
    \sum_{i=0}^{m}\binom{n}{i} > u\,.
\end{align}
Then, the differential codeword is 
\begin{align}
\label{eq:MPPMFormulation}
    \mathbf{d}_{j} = \text{MPPM}\left(u - \sum_{i=0}^{m-1}\binom{n}{i}; \, m\right)
\end{align}
where $\text{MPPM}\left(x;\,m\right)$ is the sequence of the $m$ pulses injectively associated to the input $x \in \mathbb{N}$. 
For the realization of the encoder, the binomials in \eqref{eq:findmpulses} and \eqref{eq:MPPMFormulation} can be pre-computed, so what remains to discuss is how to implement \ac{MPPM} modulators. 
%
%
%
We use the combinatorial number system for the representation of integers \cite{Dav:66,Cov:73,Knuth2011:Art}. 
It provides a one-to-one mapping between natural numbers and combinations, such that, for any integer $0 \leq x < \binom{n}{m}$, there exists a unique $m$-tuple of integers $(s_1,\dots,s_m)$, with $0 \leq s_1 < \dots < s_{m} < n$, giving 
\begin{align}\label{eq:Macaulay_u}
    x = \binom{s_1}{1} + \binom{s_2}{2} + \dots + \binom{s_m}{m} 
\end{align}
where $s_\ell$ for $\ell = m, \dots, 1$ is given by
\begin{align}\label{eq:Macaulay_s_h}
    s_\ell = \max \bigg\{ i <n : \binom{i}{\ell} \leq x - \sum_{j=\ell+1}^{m} \binom{s_j}{j} \bigg\} \,.
\end{align}
The $m$-tuple $(s_1,\dots,s_m)$ in our case gives the position of the pulses (the bit at $1$) in the $n$-tuple.
Therefore, we can use \eqref{eq:Macaulay_s_h} and \eqref{eq:Macaulay_u} to construct the $\text{MPPM}\left(x;\,m\right)$ encoder and decoder, respectively. 
The feasibility of the decoder primarily depends on the possibility to compute binomial coefficients. On the other hand, the encoder is computationally more demanding, as it requires both the computation of binomial coefficients and solving \eqref{eq:Macaulay_s_h}. 

Assuming  pre-computation and storing of the binomial coefficients, we present in Fig.~\ref{fig:OptSchemeImpl} a practical implementation able to output a codeword $\text{MPPM}\left(x; \,m\right)$ in $m$ clocks. 

%
To explain the scheme we observe that, from \eqref{eq:Macaulay_u} and \eqref{eq:Macaulay_s_h}, we need $m (n-m+1)$ $n$-bit long binomial coefficients for a \ac{MPPM} modulator using $m$ pulses and $n$ slots\footnote{It is possible to slightly reduce the number of pre-computed binomial coefficients considering only the non-zero coefficients. 
}.
The scheme iteratively compute \eqref{eq:Macaulay_u} by solving \eqref{eq:Macaulay_s_h} for $\ell = m, \dots, 1$. 
To solve \eqref{eq:Macaulay_s_h}, we make all ``less or equal'' comparisons in parallel $\forall i < n$. 
Then, a second block takes the last satisfied comparison, and creates a temporary sequence with the $s_\ell$-th bit set to ``1'' and all others to zero, while feedbacking the value $\binom{s_j}{j}$ to update the input. 
It is important to note that, since all the required $d_\mathrm{max}+1$ \ac{MPPM} modulators use $n$-tuples and observing that the binomial coefficients for $m$ pulses include, as a subset, those for a smaller number of pulses, the circuit in Fig.~\ref{fig:OptSchemeImpl} can be efficiently reused to implement the whole optimal line encoder.

The complexity of the proposed scheme is mainly in the realization of the pre-computed binomial coefficients (i.e., connections to the power supply and ground). Similarly, the decoder can first reconstruct $x$ using the same pre-computed binomial coefficients following \eqref{eq:Macaulay_u}. Then, it will compute the message $u$ as
\begin{align}
    u = x + \sum_{i=0}^{m-1}\binom{n}{i}
\end{align}
where $m$ is the number of pulses in the received message.

The general optimal scheme shown in Section~\ref{subsec:GeneralOptimalScheme} requires the computation of \eqref{eq:Macaulay_s_h}. 
Given that we need to generate a sequence with $m$ pulses, we have a cost of $nm\,\mathcal{C}_\mathrm{c}(n) + 2m\,\mathcal{C}_\mathrm{a}(n)$, as illustrated in Fig.~\ref{fig:OptSchemeImpl},
where $\mathcal{C}_\mathrm{c}(n)$ is the cost to perform a comparison between two $n$ bit sequences representing an integer value such as a ``$<$'' or a ``$\le$'', and $\mathcal{C}_\mathrm{a}(n)$ is the cost of an addition or subtraction of two $n$ long sequences representing integer values.
Averaging over all the number of possible sequences with exactly $m$ pulses, 
we have an average encoding complexity of $(n + 2) D_\mathrm{opt}(k,b)\,\mathcal{C}_\mathrm{c}(n)$ for the circuit in Fig.~\ref{fig:OptSchemeImpl}. 
To this cost we have to add the cost for deciding how many pulses are in the output sequence, which is $(d_\mathrm{max} + 1)\,\mathcal{C}_\mathrm{c}(n)$. 
The encoding complexity is 

$$\left[(n + 2) D_\mathrm{opt}(k,b)+d_\mathrm{max} + 1\right] \times \mathcal{C}_\mathrm{c}(n)
\,.$$

\begin{figure}[t]
    \centering
    \resizebox{0.8\columnwidth}{!}{
%
%
\definecolor{testGreen}{HTML}{36a447}
\definecolor{testBlue}{HTML}{3771c8}
\definecolor{testRed}{HTML}{c83737}
\definecolor{testYellow}{HTML}{ffcc00}
\definecolor{testViolet}{HTML}{834dd3}
\definecolor{testOrange}{HTML}{ea7527}

\definecolor{innerRed}{HTML}{ea4a4a}
\definecolor{innerGreen}{HTML}{48be58}

\definecolor{innerYellow}{HTML}{FFF9D2}

\begin{tikzpicture}
\begin{axis}[%
width=4.521in,
height=3.5in,
at={(0in,0in)},
scale only axis,
xmode=log,
xmin=1,
xmax=10000,
title={$k=11$},
ytick distance=0.1,
xtick = {1, 10, 100, 1e3, 1e4},
xticklabels = {$1$, $10$, $100$, $1000$, $10^4$},
xminorticks=true,
xlabel style={font=\color{white!15!black}, font=\Large},
xlabel={$b$, added lines},
ymin=0,
ymax=1,
ylabel style={font=\color{white!15!black}, font=\Large},
ytick = {0, 0.1, 0.2, 0.3, 0.4, 0.5, 0.6, 0.7, 0.8, 0.9, 1},
yticklabels={$0$, $10$, $20$, $30$, $40$, $50$, $60$, $70$, $80$, $90$, $100$},
ylabel style={font=\color{white!15!black}, font=\Large},
ylabel={Energy Saving [$\%$]},
tick label style={black, semithick, font=\Large},
axis background/.style={fill=white},
xmajorgrids,
ymajorgrids,
legend style={at={(0.97,0.03)}, anchor=south east, legend cell align=left, align=left, draw=white!15!black, font=\Large}
]
\addplot [name path=LB,color=black, line width=1.5pt]
  table[row sep=crcr]{%
0	0\\
1	0.155184659090909\\
2	0.231001420454545\\
3	0.274058948863636\\
4	0.326615767045455\\
5	0.348366477272727\\
6	0.362127130681818\\
7	0.377485795454545\\
8	0.39453125\\
9	0.413352272727273\\
10	0.434037642045455\\
11	0.456676136363636\\
12	0.481356534090909\\
13	0.483575994318182\\
14	0.485884232954545\\
15	0.48828125\\
16	0.490767045454545\\
17	0.493341619318182\\
18	0.496004971590909\\
19	0.498757102272727\\
20	0.501598011363636\\
21	0.504527698863636\\
22	0.507546164772727\\
23	0.510653409090909\\
24	0.513849431818182\\
25	0.517134232954545\\
26	0.5205078125\\
27	0.523970170454545\\
28	0.527521306818182\\
29	0.531161221590909\\
30	0.534889914772727\\
31	0.538707386363636\\
32	0.542613636363636\\
33	0.546608664772727\\
34	0.550692471590909\\
35	0.554865056818182\\
36	0.559126420454545\\
37	0.5634765625\\
38	0.567915482954545\\
39	0.572443181818182\\
40	0.577059659090909\\
41	0.581764914772727\\
42	0.586558948863636\\
43	0.591441761363636\\
44	0.596413352272727\\
45	0.601473721590909\\
46	0.606622869318182\\
47	0.611860795454545\\
48	0.6171875\\
49	0.622602982954545\\
50	0.628107244318182\\
51	0.633700284090909\\
52	0.639382102272727\\
53	0.642223011363636\\
54	0.642311789772727\\
55	0.642400568181818\\
56	0.642489346590909\\
57	0.642578125\\
58	0.642666903409091\\
59	0.642755681818182\\
60	0.642844460227273\\
61	0.642933238636364\\
62	0.643022017045455\\
63	0.643110795454545\\
64	0.643199573863636\\
65	0.643288352272727\\
66	0.643377130681818\\
67	0.643465909090909\\
68	0.6435546875\\
69	0.643643465909091\\
70	0.643732244318182\\
71	0.643821022727273\\
72	0.643909801136364\\
73	0.643998579545455\\
74	0.644087357954545\\
75	0.644176136363636\\
76	0.644264914772727\\
77	0.644353693181818\\
78	0.644442471590909\\
79	0.64453125\\
80	0.644620028409091\\
81	0.644708806818182\\
82	0.644797585227273\\
83	0.644886363636364\\
84	0.644975142045455\\
85	0.645063920454545\\
86	0.645152698863636\\
87	0.645241477272727\\
88	0.645330255681818\\
89	0.645419034090909\\
90	0.6455078125\\
91	0.645596590909091\\
92	0.645685369318182\\
93	0.645774147727273\\
94	0.645862926136364\\
95	0.645951704545455\\
96	0.646040482954545\\
97	0.646129261363636\\
98	0.646218039772727\\
99	0.646306818181818\\
100	0.646395596590909\\
125	0.648615056818182\\
150	0.650834517045455\\
175	0.653053977272727\\
200	0.6552734375\\
225	0.657492897727273\\
250	0.659712357954545\\
275	0.661931818181818\\
300	0.664151278409091\\
325	0.666370738636364\\
350	0.668590198863636\\
375	0.670809659090909\\
400	0.673029119318182\\
425	0.675248579545455\\
450	0.677468039772727\\
475	0.6796875\\
500	0.681906960227273\\
525	0.684126420454545\\
550	0.686345880681818\\
575	0.688565340909091\\
600	0.690784801136364\\
625	0.693004261363636\\
650	0.695223721590909\\
675	0.697443181818182\\
700	0.699662642045455\\
725	0.701882102272727\\
750	0.7041015625\\
775	0.706321022727273\\
800	0.708540482954545\\
825	0.710759943181818\\
850	0.712979403409091\\
875	0.715198863636364\\
900	0.717418323863636\\
925	0.719637784090909\\
950	0.721857244318182\\
975	0.724076704545455\\
1000	0.726296164772727\\
1250	0.748490767045455\\
1500	0.770685369318182\\
1750	0.792879971590909\\
2000	0.815074573863636\\
2036	0.818270596590909\\
3000    0.818270596590909\\
4000    0.818270596590909\\
5000    0.818270596590909\\
6000    0.818270596590909\\
7000    0.818270596590909\\
8000    0.818270596590909\\
9000    0.818270596590909\\
10000    0.818270596590909\\
};
\addlegendentry{Optimal}

\addplot [name path =asseX,color=black, draw opacity=0, dashed, line width=1.2pt,forget plot]
  table[row sep=crcr]{%
1	      1\\
10000     1\\
};
\addplot [thick, color=innerYellow, fill=innerYellow, fill opacity=0.5,forget plot]
fill between[
    of = LB and asseX,
    soft clip = {domain=1:1e4},
];

\addplot [color=black, dashed, line width=1.7pt]
  table[row sep=crcr]{%
1	    0.818270596590909\\
10000	0.818270596590909\\
};
\addlegendentry{PPM Bound}

\addplot [color=testBlue, line width=2pt, mark size=4pt, mark=triangle*, mark options={solid, fill=brightBlue}]
  table[row sep=crcr]{%
1	0.1552\\
};
\addlegendentry{Bus Inversion}

\addplot [color=testRed, line width=2pt, mark size=3.5pt, mark=*, mark options={solid, fill=innerRed}]
  table[row sep=crcr]{%
12	0.4814\\
};
\addlegendentry{Syndrome-based Golay}

\addplot [color=testGreen, line width=2pt, mark size=3.7pt, mark=square*, mark options={solid, fill=innerGreen}]
  table[row sep=crcr]{%
2047	0.818270596590909\\
};
\addlegendentry{PPM$_0$}

\node[right] (A) at (axis cs:1.5,0.65) {\fcolorbox{black}{white}{\large Forbidden Region}};

\end{axis}


\end{tikzpicture}%
    }
    \caption{Energy saving $1-D_\mathrm{opt}(k,b)/D_\mathrm{unc}(k)$ for the Optimal Bus Encoding Scheme, information size $k=11$.}
 \label{fig:optimum_k11}
\end{figure}
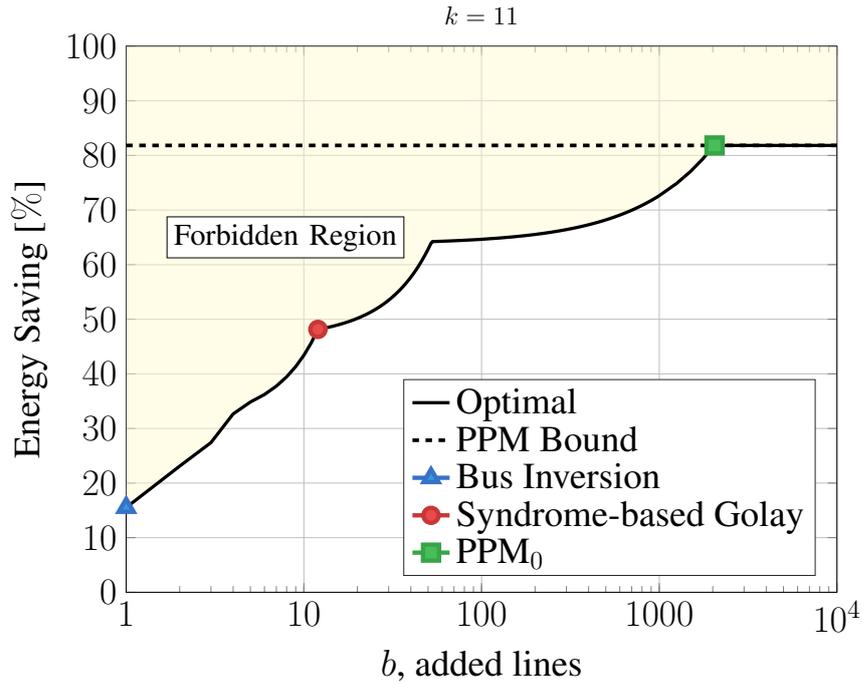

The optimal code performance as derived from \eqref{eq:Dopt} for $k=11$ is reported in Fig.~\ref{fig:optimum_k11} for different values of the redundancy $b$. 



\bibliographystyle{IEEEtran}
\bibliography{Files/IEEEabrv,Files/StringDefinitions,Files/StringDefinitions2,Files/refs}



\end{document}